%% file: secpm-arXiv.tex
\documentclass[letterpaper,twocolumn,10pt]{article}
\usepackage{usenix2019,epsfig,endnotes}

\usepackage{booktabs}

\usepackage{marvosym}

\usepackage{array}

\usepackage{graphicx}
\usepackage{epstopdf}
\usepackage{comment}

\usepackage{amsmath}
\usepackage{arydshln}
\usepackage{algorithmic}
\renewcommand{\algorithmiccomment}[1]{\bgroup\hfill//~#1\egroup}

\usepackage[linesnumbered,ruled]{algorithm2e}
\usepackage[shortlabels]{enumitem}
\usepackage{subfigure}
\usepackage{multirow}
\usepackage{multicol}

\usepackage{url}

\usepackage{color}
\definecolor{MyGreen}{RGB}{0,100,0}
\definecolor{MyRed}{RGB}{255,0,102}

\usepackage[colorlinks=true, linkcolor=red, anchorcolor=red, citecolor=red]{hyperref}

\usepackage{lipsum}
\usepackage{titlesec}
\usepackage{makecell}

\titlespacing\section{0pt}{7pt plus 4pt minus 2pt}{5pt plus 2pt minus 2pt}
\titlespacing\subsection{0pt}{7pt plus 3pt minus 2pt}{5pt plus 2pt minus 2pt}
\titlespacing\subsubsection{0pt}{7pt plus 3pt minus 2pt}{6pt plus 2pt minus 2pt}

\begin{document}

\hyphenpenalty = 8000
\tolerance = 2000

\date{}

\title{\Large \bf A Scalable Learned Index Scheme in Storage Systems~\vspace{-0.4cm}}

\author{
{\rm Pengfei Li, Yu Hua, Pengfei Zuo, Jingnan Jia
}\\
Huazhong University of Science and Technology \\
}

\maketitle


\subsection*{Abstract}
\noindent
Index structures are important for efficient data access, which have been widely used to improve the performance in many in-memory systems. Due to high in-memory overheads, traditional index structures become difficult to process the explosive growth of data, let alone providing low latency and high throughput performance with limited system resources. The promising learned indexes leverage deep-learning models to complement existing index structures and obtain significant memory savings. However, the learned indexes fail to become scalable due to the heavy inter-model dependency and expensive retraining. To address these problems, we propose a scalable learned index scheme to construct different linear regression models according to the data distribution. Moreover, the used models are independent so as to reduce the complexity of retraining and become easy to partition and store the data into different pages, blocks or distributed systems. Our experimental results show that compared with state-of-the-art schemes, AIDEL improves the insertion performance by about 2$\times$ and provides comparable lookup performance, while efficiently supporting scalability.

\input{introduction.tex}
\input{background.tex}
\input{design.tex}

\input{experiment.tex}

\input{related_work.tex}

\section{Conclusion}
In order to address the problem of scalability of learned indexes, we present a scalable learned index scheme, called AIDEL. In the context of our paper, the models are interpreted as regression ML models with bounded prediction errors, which are used to predict the positions of the keys by learning the data distribution. Unlike existing learned indexes, our models are independent in inter- or inner-layers, which are generated by our LPA algorithm. The LPA algorithm eliminates invalid and redundant models in the learned indexes, since this algorithm adaptively assigns different regression models according to the data distribution. The independency between models enables AIDEL to partition the data and store them into different regions with low overheads in terms of retraining. AIDEL handles inserts through a structure of sorted lists, which keep all ordered data to efficiently meet range requests. Unlike the traditional tree-based indexes, the searching process in our scheme is mainly achieved by calculations. Our experimental results show that compared with the B$^+$-tree, AIDEL improves 1.3$\times$ to 2.7$\times$ insertion throughput and about 2$\times$ lookup throughput. Compared with learned indexes, AIDEL provides comparable lookup performance while efficiently supporting scalability.

{
\bibliographystyle{acm}
\bibliography{reference}}


\end{document}

%% file: introduction.tex
\section{Introduction}
Efficient data storage and access are important for both industry and academia, and the explosive growth of data exacerbates this problem. Index structures, such as B$^+$-tree, Hash-map, and Bloom filters usually support today's in-memory systems to handle data processing tasks according to different requirements~\cite{alexiou2013adaptive,fan2014cuckoo,graefe2001b,richter2015seven}. Traditional index structures have been improved to be more memory-efficient over the past decades~\cite{graefe2001b, kim2010fast, rao1999cache, chang2000b}.

Tree-based structures keep all ordered data for range requests, which aim to identify the items within a given range. A common approach to build a low-latency and high-throughput storage system is to maintain all data and metadata completely in the main memory, which eliminates the expensive disk I/O operations~\cite{wu2018wormhole}. In fact, the indexes, e.g., tree-based structures, consume around 55\% of the total memory in a state-of-the-art in-memory systems~\cite{zhang2016reducing}. The expensive space overhead becomes exacerbated when the index structures are too large to fit into the limited-size memory. 

Existing works attempt to improve the performance and reduce storage overhead. For example, since the B$^+$-tree exhibits good cache-line locality, many cache-conscious variants of B$^+$-tree including CSS-tree~\cite{rao1999cache}, CSB-tree~\cite{rao2000making} and FAST~\cite{kim2010fast} have been developed. Some schemes also propose to use hybrid index structures to further improve the performance via GPUs~\cite{kaczmarski2012b+,kim2010fast,shahvarani2016hybrid}. Moreover, in order to reduce memory overhead of the B$^+$-tree, compression schemes, including prefix/suffix truncation, dictionary compression and key normalization have been proposed~\cite{goldstein1998compressing,bayer1977prefix,boehm2011efficient,rao1999cache,neumann2008rdf,zukowski2006super}. Some schemes use approximate structures to process the indexes~\cite{athanassoulis2014bf,galakatos2018tree,kraska2018case}.

However, all above schemes are designed for general-purpose data structures and mainly focus on the index structures themselves, thus overlooking the patterns of data distribution. Kraska et al~\cite{kraska2018case} argue that exact data distribution enables high optimization for almost any index structure. For example, a linear regression function is sufficient for a system to store and access a set of continuous integer keys (e.g., the keys from 1 to 100M), which has significant advantages over traditional B$^+$-trees in terms of lookup performance and memory overhead. The patterns of data distribution become important for storage systems to deliver high performance. However, in real-world applications (e.g., web servers), it's usually difficult to accurately obtain the patterns of data distribution in advance and some patterns may be extremely complex or even impossible to represent via known patterns. We hence consider machine learning (ML) approaches to learn a model that exhibits the patterns of data distribution, called \textbf{\textit{learned indexes}}~\cite{kraska2018case}.

In the context of our paper, the models are interpreted as regression ML models with bounded prediction errors, which are used to predict the positions of the keys by learning the data distribution. However, the data generated from the real-word applications, e.g., web servers, Internet of Things (IoT), autonomous vehicles, are extremely hard to learn. Inspired by the learned indexes, we partition the data into different parts and use multiple linear regression models to learn each part well~\cite{kraska2018case}. In order to meet range requests, all data are sorted for further training. Unlike the traditional tree-based indexes, the searching process in learned indexes is mainly achieved by calculations. For example, the trained linear regression models offer a prediction range based on the queried key, and guarantee that the prediction range contains the key if existing.

The learned indexes open up a new perspective on indexing issues: \textbf{\textit{indexes can be considered as ML models}}. We use cost-efficient computation to speed up traditional comparison operations, thereby increasing access speed and saving memory resources. However, it is non-trivial to efficiently leverage learned indexes due to the following challenges.
\begin{enumerate}[nosep, leftmargin=0pt, itemindent=3em]
	\item[\textbf{\textit{1)}}] \textbf{\textit{Poor Scalability.}} The poor scalability comes from the heavy inter-model dependency and expensive retraining. Specifically, to keep all ordered data for range requests, inserting new data into learned indexes will change the positions of some data, thus leading to many data movements or even increasing the probabilities that some data can't be found. To ensure the accuracy of the models, all the models have to be retrained even if only one model needs to be updated, since all the models are highly dependent. Furthermore, it is hard for learned indexes to partition and store the data into different regions, since learned indexes assume that all the data are stored in one continuous block, and we can't move any models unless retraining.
	\item[\textbf{\textit{2)}}] \textbf{\textit{High Overheads.}} Learned indexes build a delta-index~\cite{severance1976differential} to handle inserts, which however produces high overhead. For example, the index structure needs to be retrained when the delta-buffer is full. Moreover, handling range requests is inefficient, since the data are stored in different structures and hence all the data are not in order. Learned indexes argue to leverage the recursive model index structure~\cite{kraska2018case} or build an additional translation table to partition the data, but the two schemes are inefficient and cause extra costs. Such designs require extra space and need to be rebuilt during retraining.
\end{enumerate}

In order to address these challenges, our paper presents a scalable and adaptive learned index scheme with a bounded prediction error, called AIDEL. Unlike existing learned indexes, our models are completely independent in inter- or inner-layers. All models are generated through our main component, Learning Probe Algorithm (LPA), which adaptively assigns linear regression models according to the data distribution. AIDEL achieves scalability through the sorted lists, which can easily handle inserts and keep all ordered data to efficiently meet range requests. Such designs in AIDEL allow to update any models without affecting the entire structure.

The key contributions of this paper are summarized.
\begin{itemize}[nosep, leftmargin=0pt, itemindent=3em]
	\setlength{\itemsep}{0pt}
	\setlength{\parsep}{0pt}
	\setlength{\parskip}{0pt}
	\item \textbf{\textit{High Scalibility.}} We present a scalable learned index scheme, AIDEL, which eliminates the dependency of models and handles inserts through sorted lists. Unlike existing learned indexes, AIDEL partitions the data and stores them into different regions with low overheads in terms of retraining, thus efficiently supporting system scalability.
	\item \textbf{\textit{Strong Adaptivity.}} We propose learning probe algorithm to adaptively assign different linear regression models according to the data distribution. This algorithm not only ensures all the models are independent, but also reduces the number of models to save more space than learned indexes. Our scheme hence obtains adaptivity to efficiently handle various requests.
	\item \textbf{\textit{Low Overheads.}} Retraining and updating AIDEL are cost-efficient, since we can retrain and update any models without affecting the entire structure due to the independency between models. AIDEL not only efficiently meets range requests, but also saves time on re-sorting the data during retraining, since AIDEL ensures all the data are kept in order, while also significantly reducing the overheads.
\end{itemize}

The rest of this paper is organized as below. Section 2 introduces the background. In Section 3, we present the learning probe algorithm, and then demonstrate the idea and main operations of AIDEL. Section 4 shows the experimental results and analysis. Section 5 discusses the related work, and Section 6 concludes our paper.

%% file: background.tex
\section{Background and Motivation}

\subsection{Index Structures for Range Requests}\label{section:range_index}
In general, an index structure is able to support point query that searches for a given item. Unlike it, a range query aims to identify the items within a given range, which requires the data to be sorted, thus facilitating efficient data accessing. Due to the salient features of efficiency and scalability, B$^+$-tree~\cite{comer1979ubiquitous} is able to meet the needs of requirements of real-world applications.

First, B$^+$-tree is efficient for range requests. B$^+$-tree stores all data in the leaf nodes and keeps the data in order, which enables efficient range requests. In order to find the queried data, inner nodes are used to indicate which nodes to be accessed next, until the leaf nodes are found. During the searching, all the data in a node will be accessed if this node is selected by the previous one. Rao el al~\cite{rao1999cache} showed that the B$^+$-tree exhibits good cache behaviors, since all the data in a node are accessed and used in more comparisons by one cache line if the length of a node is aligned with the cache line. Thus, CSS-tree~\cite{rao1999cache} and CSB+-tree~\cite{rao2000making} are proposed to provide efficient lookup performance by exploiting the cache. Recently, FAST~\cite{kim2010fast} tries to make use of SIMD to further improve the performance. However, these optimized B$^+$-trees need to allocate more memory for the inner nodes. The nodes need to be realigned with cache and SIMD when new insertions occur.

Second, B$^+$-tree achieves scalability by dynamically balancing the tree size. New data can be inserted if the found leaf node is not full, and otherwise more empty positions will be generated by splitting and merging the nodes. Based on the dynamic-size feature, B$^+$-tree can easily handle inserts and keep all ordered data to efficiently support range requests~\cite{hwang2018endurable}. However, the size of B$^+$-tree keeps growing with the growth of the inserts and inner nodes consume a significant amount of available memory, which dramatically decrease the lookup performance once the index structure overfolws the memory.

Third, B$^+$-tree provides correctness guarantee that the data are promised to be found once inserted. This correctness guarantee seems to be a fundamental feature of the index structures, which however is not easy to be satisfied by learned indexes, especially for the newly inserted data. Because the inserted data will change the positions of some data, leading to a failure probability that some data can't be found since their new positions exceed the predicted range. More details are analyzed in Section \ref{section:scalability}.

\subsection{New Perspective on Index Issues}\label{section:new_persepctive}
\begin{figure}[htbp]
	\centering\includegraphics[width=3.5in]{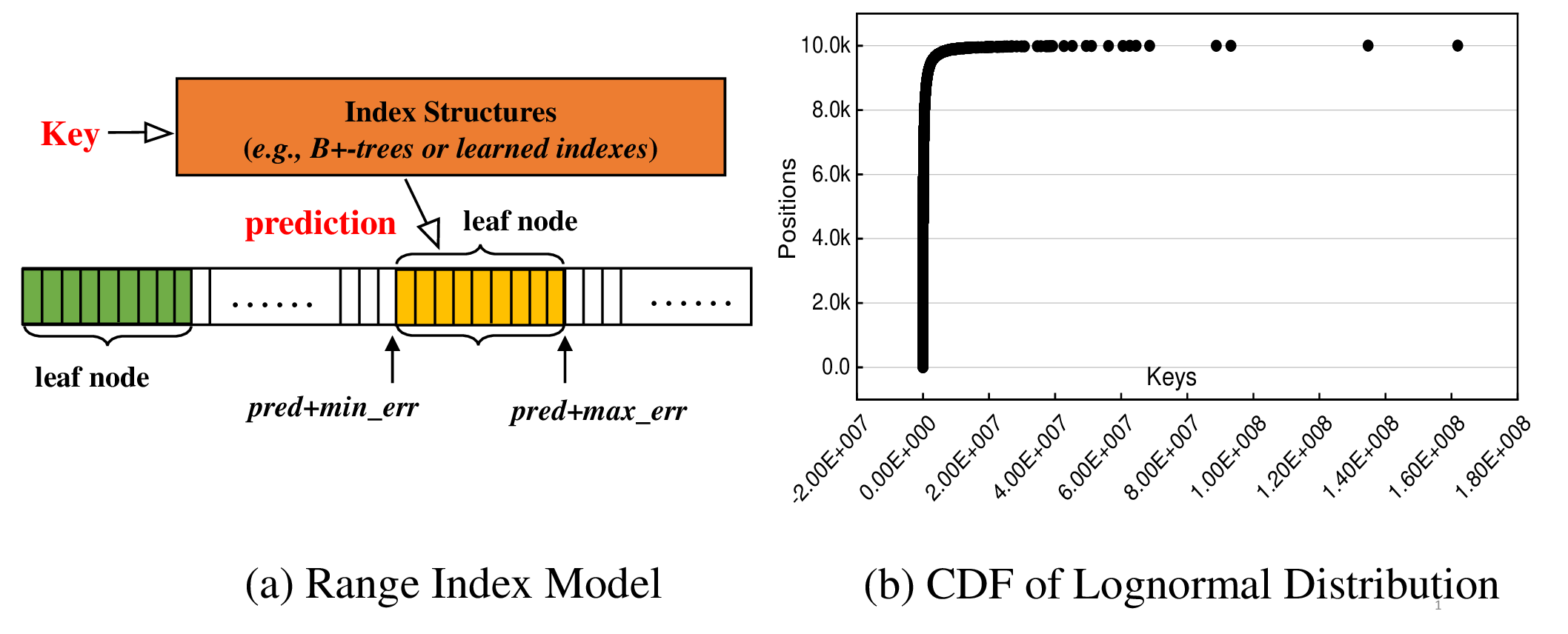}
	\caption{Range index and CDF models.}\label{fig:index_model}
\end{figure}

From the perspective of machine learning, range index structures are regression models~\cite{kraska2018case}, which can predict the position of a given look-up key as shown in Figure \ref{fig:index_model}(a). The index structures in Figure \ref{fig:index_model}(a) can be the B$^+$-tree or learned indexes. In the B$^+$tree, the data are stored in leaf nodes and can be found through checking the tree. Learned indexes~\cite{kraska2018case} view this process as a prediction, and the records between $[pred+min\_err, pred+max\_err]$ (where $min\_err$ may be a negative) can be considered as the same concept with the leaf nodes in the B$^+$-tree. Obviously, the length of $[pred+min\_err, pred+max\_err]$ will affect the lookup performance, and we term this length as \textbf{\textit{prediction granularity}}.

For making the prediction practical, the sorted keys and the true positions are considered as inputs and outputs, respectively. The relationship between keys and positions is similar to a cumulative distribution function (CDF) as shown in Figure \ref{fig:index_model}(b). The dataset used in Figure \ref{fig:index_model}(b) is the same as that used in Section \ref{section:experiment}, which is synthesized by lognormal distribution with $\mu$=0 and $\sigma$=2. Based on this observation, the prediction accuracy can be improved by learning the patterns of data distribution according to the CDF. Numerous schemes have been proposed to estimate the distribution of data~\cite{dvoretzky1956asymptotic,magdon1999neural,huang2011cumulative}, which can be used in our work.

When the CDF between keys and positions can be accurately represented via the known regression models, the lookup complexity is $O(1)$ since each position can be calculated by the regression models. For example, a set of continuous integer keys (e.g., the keys from 1 to 100M) are stored in a piece of continuous positions (e.g., positions from 1 to 100M). The CDF can be represented as $y=x$, where $x$ and $y$ are keys and positions respectively. The prediction granularity is 1, which means that this regression model is accurate enough without any prediction error. However, in real-world applications (e.g., web servers), the CDFs can't be obtained in advance and some CDFs may be extremely complex or even impossible to represent via known regression models~\cite{kraska2018case}. In these situations, we don't need to accurately represent the CDF to reduce the prediction granularity to 1, since the length of the leaf nodes in the B$^+$-tree has never been set to 1, which simplifies the prediction problem: regression models only need to approximately represent the CDF and reduce the prediction granularity to the same size like the leaf nodes in the B$^+$-tree.

In this paper, we evaluate several different methods to approximately represent the CDF and the results are shown in Figure \ref{fig:approximate}. Our proposed LPA algorithm learns the CDF best with the same number of models, which will be elaborated in Section \ref{section:lepa_algorithm}.

\subsection{Learned Indexes}\label{section:learned_indexes}
One of the key challenges of using the learned model as an index structure is how to provide a small prediction granularity to find the lookup key quickly as a B$^+$-tree does. Using a single ML model to reduce the prediction granularity (e.g., from 100M to 10) is difficult, which may result in an extremely complex ML model. It is hard to design and train this type of models. However, it is easy to reduce the prediction granularity from 100M to 10K, then from 10K to 100, via various small ML models. Based on this observation and inspired by the mixture of experts work (a type of ML models for complex tasks)~\cite{shazeer2017outrageously}, learned indexes propose to use a recursive model index (RMI) to improve the prediction accuracy.

\begin{figure}[tbp]
	\centering\includegraphics[width=3.3in]{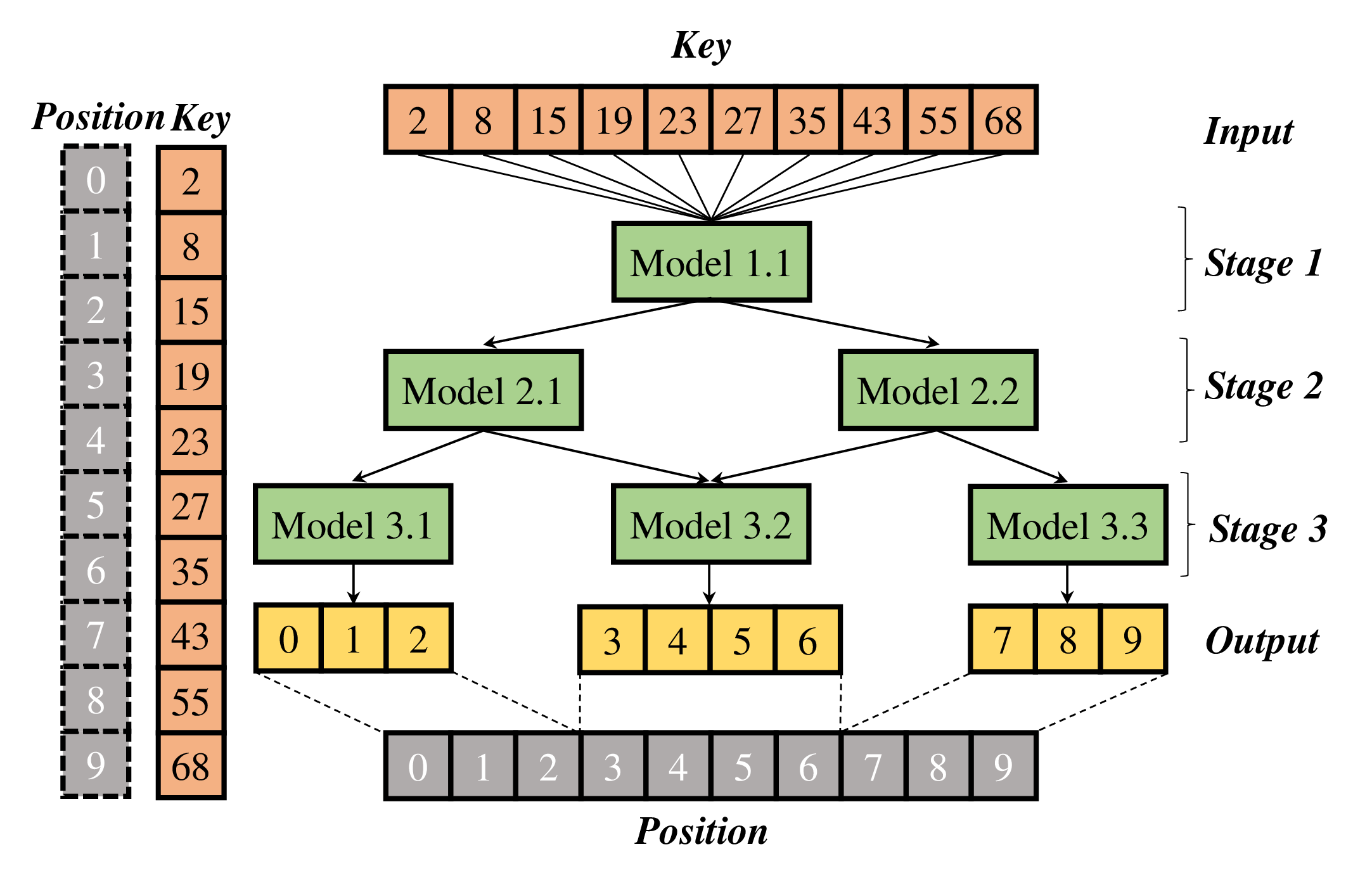}
	\caption{The RMI in learned indexes to improve the prediction accuracy.}\label{fig:learned_index}
\end{figure}

The main idea of RMI is to build a hierarchy of models where at each stage the model picks another model based on the intermediate prediction results, until the final stage predicts the position~\cite{kraska2018case}. As shown in Figure \ref{fig:learned_index}, a RMI consists of 3 stages, respectively containing 1, 2, 3 ML models. These models are trained in the order of hierarchical relationship, each of which is trained with different training data. For example, Model 1.1 in the top level is trained first with the whole dataset that contains $N$ entries as shown in Figure \ref{fig:learned_index}. Based on the prediction results of Model 1.1, either Model 2.1 or 2.2 is selected and the entire dataset is also divided into two subdatasets according to the selection results. Then, the two models in the second stage are trained with their respective subdatasets. The next stage follows the similar training process.

In order to provide the correctness guarantee that learned indexes can accurately find the queried key, learned indexes store the min- and max-error for every model in the last stage, which can be calculated as follows:
\begin{equation}\label{equation:calculate_error}
\begin{aligned} 
min\_err=min(y_{i}-f_{L}^{j}(x))\quad \forall i\in S_{L.j},j\in M_{L} \\
max\_err=max(y_{i}-f_{L}^{j}(x))\quad \forall i\in S_{L.j},j\in M_{L}
\end{aligned}
\end{equation}
where $y_{i}$ represents the true position of each key in the subdataset $S_{L.j}$, $f_{L}^{j}(x)$ represents the prediction result of $j_{th}$ model in the last stage L and there are $M_{l}$ models in stage $l$. If absolute min-/max-error is above the predefined threshold, the ML model becomes invalid to be replaced with a B$^+$-tree. Finally, learned indexes show the prediction granularity $[pred+min\_err, pred+max\_err]$ if the picked ML model is valid, and otherwise the lookup key will be searched in the B$^+$-tree.

Learned indexes implement a 2-stage RMI index with a small neural network (NN) on the top and a large amount of linear regression models at the bottom. Because learned indexes observe that a simple (0 hidden layers) to semi-complex NN model (2 hidden layers) on the top works better than other configured NN models (i.e., with more hidden layers). It is not cost-efficient to execute complex models at the bottom since the simple linear regression models are accurate enough to learn the small subdatasets.

Our experiments demonstrate that there exists invalid and redundant models in learned indexes, as shown in Sections \ref{section:strategy_comparison} and \ref{section:model_numbers}. Specifically, RMI needs to be configured in advance (e.g., ML models, threshold of error, etc.), the models with smaller prediction errors than the threshold are considered to be valid, and otherwise traditional index structures have to be used. However, the prediction errors of some models will exceed the threshold if we don't configure enough models, causing these models to be invalid. Simply increasing the number of models to eliminate invalid models will result in many redundant models. Because the prediction errors of valid models have been smaller than the threshold, leading to the newly increased models to be redundant. The main reason for generating invalid and redundant models is that learned indexes can't dynamically allocate models according to data distribution.

\subsection{Performance Guarantee}\label{section:competitive_performance}
One concern about using ML models for indexing is the calculation overhead because ML models are used to handle complex tasks including image recognition, natural language processing, robotics, etc. They are usually considered computation-intensive and storage-intensive due to heavy computation consumption in training and inference~\cite{krizhevsky2012imagenet,simonyan2014very}.

However, in learned indexes, finite arithmetic operations are faster than that traveling a B$^+$-tree even if the B$^+$-tree is in the cache~\cite{kraska2018case}. In this paper, AIDEL only contains linear regression models whose arithmetic operations are simple. In the meantime, besides the wide use of Graphics Processing Unit (GPU) and Tensor Processing Unit (TPU), there are also many researches on machine learning accelerators including GPUs, FPGAs, ASICs, PIMs and NVMs~\cite{chen2014diannao,chen2016eyeriss,kim2016neurocube,wang2017towards,chi2016prime,shafiee2016isaac}.

It is worth noting that the design goal of AIDEL is not to completely replace the traditional index structures. AIDEL is orthogonal to the traditional index optimization methods such as compression techniques and cache-conscious approaches~\cite{rao1999cache, rao2000making, kim2010fast, goldstein1998compressing,bayer1977prefix,boehm2011efficient,rao1999cache,neumann2008rdf,zukowski2006super}.

%% file: design.tex
\section{The AIDEL Design}
In this section, we elaborate the design of Adaptive InDEpendent Linear regression models (AIDEL) for scalable learned indexes, which can learn the CDF of data distribution.

The design goal of AIDEL is to eliminate the dependency between models in learned indexes, while providing efficient scalability. One of the key insights is to train linear regression models according to data distribution, which eliminates the invalid and redundant models. Another insight is to handle inserts via sorted lists, which not only avoids the probability that some data can't be found but also efficiently meets range requests.

The AIDEL consists of two stages. The first stage are key-value pairs which are used to indicate which model is chosen according to the lookup key, where the key is the first data covered by each model and the value is a pointer to the model. The second stage are linear regression models which are used to predict the positions. All models are generated by learning probe algorithm to eliminate the invalid and redundant models.

The reason for using $<key,model>$ pairs in the first stage is that such design eliminates the dependency between models. We can hence retrain or move any models without affecting the entire structure. The second stage uses linear regression models instead of complex neural networks, because the whole CDF can be represented via a large amount of models and the linear regression models are enough to learn each part well~\cite{kraska2018case}. Moreover, linear regression models have few parameters, which are easy to be trained and more space-efficient.

AIDEl achieves scalability through sorted lists which are appended behind existing data. The new data can be inserted into sorted lists without changing the positions of the data that have been learned by the model. Such design not only avoids the probability that some data fail to be found due to data movements but also keeps all ordered data for range requests.

\subsection{Learning Strategies}\label{section:strategy_comparison}
In the learned indexes, the lookup performance depends largely on the size of prediction granularity. To bound the worst-case performance of learned indexes to that of the B$^+$-tree, the invalid model whose error is larger than the predefined threshold is replaced with a B$^+$-tree~\cite{kraska2018case}. One way to improve the lookup performance is to eliminate the invalid models, which can be achieved by learning the CDF of data distribution to improve the prediction accuracy as analyzed in Section \ref{section:new_persepctive}.

\begin{figure}[tbp]
	\centering\includegraphics[width=0.45\textwidth]{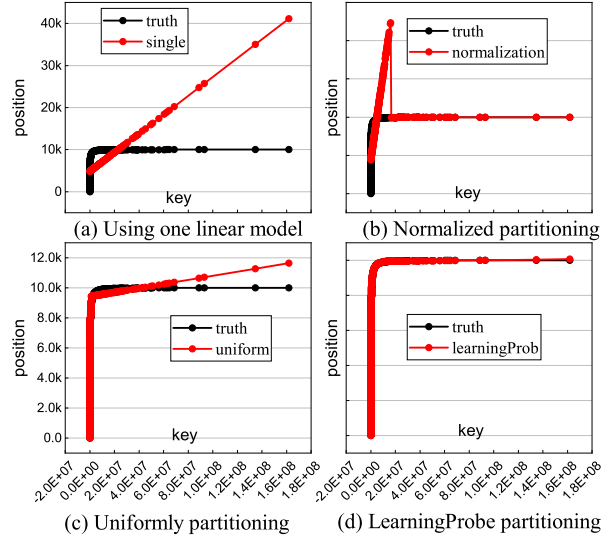}
	\caption{The efficiency of AIDEL in learning CDF.}\label{fig:approximate}
\end{figure}

We use several different strategies to learn the CDF in Figure \ref{fig:index_model}(b) and the results are shown in Figure \ref{fig:approximate}. We observe that it's impossible to represent the lognormal distribution perfectly by only using one regression model as shown of the red line in Figure \ref{fig:approximate}(a). Because the distributions in real-world applications are more complex than linear distribution~\cite{ galakatos2018tree, kraska2018case}. According to the idea of learned indexes that we can approximately represent the data distribution by dividing the data into different parts and representing them with different regression models, we examine other strategies to learn the CDF.

In order to implement a 2-stage RMI, we use a linear regression model in the first stage since learned indexes identify that a simple neural network (with hidden layers from 0 to 2) for the first stage works well. A zero hidden-layer NN is equivalent to a linear regression model. We only configure 10 linear regression models in the second stage because the used dataset in this evaluation only contains 10K records and 10 models are enough to show the strengths of different strategies. Our experiments demonstrate that the way to select the models affects the prediction accuracy. Because this selection process also determines how to divide the dataset as analyzed in Section \ref{section:learned_indexes}. The linear regression models in the second stage can't fit the data distribution well if the obtained subdatasets have poor linear patterns.

We first examine the selection process of learn indexes. Formally, each ML model can be essentially treated as a mathematical function $f(x)$, in which $x$ is the queried key. If we use $f_{l}(x)$ to denote ML models in different stages, the selection process can be described as follows:
\begin{equation}\label{equation:calculation}
f_{l}(x)=f_{l}^{(\lfloor M_{l}f_{l-1}(x)/N\rfloor)}(x)\qquad f_1(x)=y
\end{equation}
where $x$ represents input, $N$ represents the total positions of the stage, $y\in(0, M_{2}]$ represents prediction result of the first model, and there are $M_{l}$ models in stage $l$. From this formulation, we find that the core idea to select the next model and partition the dataset is normalization, represented as $\lfloor M_{l}f_{l-1}(x)/N\rfloor$.

We examine the approach (i.e., normalization) of learned indexes to select the next model and the results are shown in Figure \ref{fig:approximate}(b). The matching effect of each model varies significantly depending on the data distribution. For example, the densely distributed data are not well learned while it's much better for the sparse part. The reason is that densely distributed data are likely to be divided into the same subdataset according to Equation \ref{equation:calculation}, even if these data are not linearly distributed, resulting in poor learning accuracy. Increasing the number of models allows these densely distributed data to be partitioned into multiple subdatasets, thus allowing more models to be used to improve the learning accuracy. However, the strategy of adding models will also be applied to the sparsely distributed data, while resulting in some models to be redundant. Because the sparsely distributed data have been well learned and there are no needs to increase models for this part. Thus, the number of models in learned indexes is not optimal. Moreover, we have no priori knowledge of the data distribution in advance, which increases the difficulty for configuring the number of models.

We use the same configuration as the implemented 2-stage RMI, except modifying the selection strategy between the models. In this strategy, we divide the dataset evenly so that each subdataset has the same amount of data and the results are shown in Figure \ref{fig:approximate}(c). This strategy improves the learning accuracy for densely distributed data since these data are divided into multiple subdatasets and can be learned by more models. However, this strategy reduces the learning accuracy for sparsely distributed data, since we have to add some data from densely distributed data into sparse part to achieve the same amount of data, even if these data are not linearly distributed.

Neither of these two strategies can learn CDF well. The main reason is that the two methods can't adaptively configure models based on the data distribution, which motivates us to propose the learning probe algorithm (LPA). For fair comparisons, we also use 10 models and the results are shown in Figure \ref{fig:approximate}(d). The proposed learning probe algorithm learns the CDF better than the previous strategies. Because the dataset is divided according to the data distribution, only the linearly distributed data will be divided into the same subdatast which are easy to be learned by a linear regression model. The details of this algorithm are described in Section \ref{section:lepa_algorithm}.

\subsection{The Learning Probe Algorithm}\label{section:lepa_algorithm}
To overcome the shortcomings of previous strategies, this paper proposes the learning probe algorithm (LPA), which uses the greedy strategy to adaptively partition the data according to the data distribution. Unlike existing work, in LPA, only the data with the same linear distribution can be divided into the same subdataset. Therefore, each subdataset can be easily represented by a linear regression model. The criterion for judging whether the data have the same distribution is whether the error of the obtained model exceeds a predefined threshold. LPA will add more data to the subdataset if the error of obtained model is smaller than the threshold. Otherwise, we remove a small amount of data in the order from back to front, since reducing the amount of data can decrease the prediction error of the regression model. The complete process of LPA algorithm is shown in Algorithm \ref{algorithm:lepa}.

\begin{algorithm}[tbp]
	\caption{LPA Algorithm}\label{algorithm:lepa}
	\begin{small}
		\KwIn{int $threshold$,int $learning\_step$,float $learning\_rate$,dataType $record[N]$}
		\KwOut{trained $AIDEL$}
		\While{not reach the end of the dataset record[N]}
		{
			add $learning\_step$ data into dataset $S$ from $record$\;
			train a linear regression $model$ on $S$\;
			$error$ = max$(|min\_error|, |max\_error|)$\;
			\While{$error<threshold$}
			{
				add next $learning\_step$ data into dataset $S$ from $record$\;
				train a new $model$ on $S$\;
			}
			\While{$error>threshold$}
			{
				$step$=int$(learning\_step*learning\_rate)$\;
				remove $step$ data from the end of dataset $S$\;
				train a new $model$ on $S$\;
			}
			$AIDEL$.append($model$)\;
			clean data from dataset $S$ for next probing\;
		}
	\end{small}	
\end{algorithm}

Before using LPA, we need to configure some parameters including the $threshold$, $learning\_step$ and $learning\_rate$, where $threshold$ is the max $error$ of the model we can tolerate, $learning\_step$ and $learning\_rate$ are used to determine the learning speed. As shown in Algorithm \ref{algorithm:lepa}, the main component of LPA works like a probe, which first walks forward for a large step of length $learning\_step$, i.e., add $learning\_step$ data from the training dataset $record$ into a small dataset $S$ (line 2). We can obtain a linear regression model on dataset $S$ and calculate the prediction $error$ of the model (lines 3 and 4), where $min\_err$ and $max\_err$ can be calculated by Equation \ref{equation:calculate_error}. The prediction error of the obtained model determines the next operation of the probe. If $error<threshold$, the probe keeps moving forward to another $learning\_step$ to obtain a new model until the error of obtained model is not smaller than $threshold$ (lines 5-8). When $error>threshold$, the probe keeps moving backward with a smaller step until the prediction error of the obtained model is not larger than $threshold$ (lines 9-13). Finally, LPA appends the model to AIDEL and cleans the dataset $S$ for next probing (lines 14 and 15).

Compared with learned indexes, all the models generated by LPA are valid since only the model whose prediction error is not larger than $threshold$ can be appended to AIDEL. And the max error of each obtained model can be controlled by the predefined parameter $threshold$. Benefit from the greedy strategy, LPA eliminates redundant models at the same time. Because each model covers as many continuous data as possible, where these data have the same linear distribution. Furthermore, we don't need to configure the number of models in advance, even if we have no priori knowledge of the data distribution. Because LPA can adaptively partition the data according to the data distribution. By eliminating invalid and redundant models, the number of models is far smaller than that of learned indexes, as shown in Section \ref{section:model_numbers}.

\begin{figure}[tbp]
	\centering\includegraphics[width=0.45\textwidth]{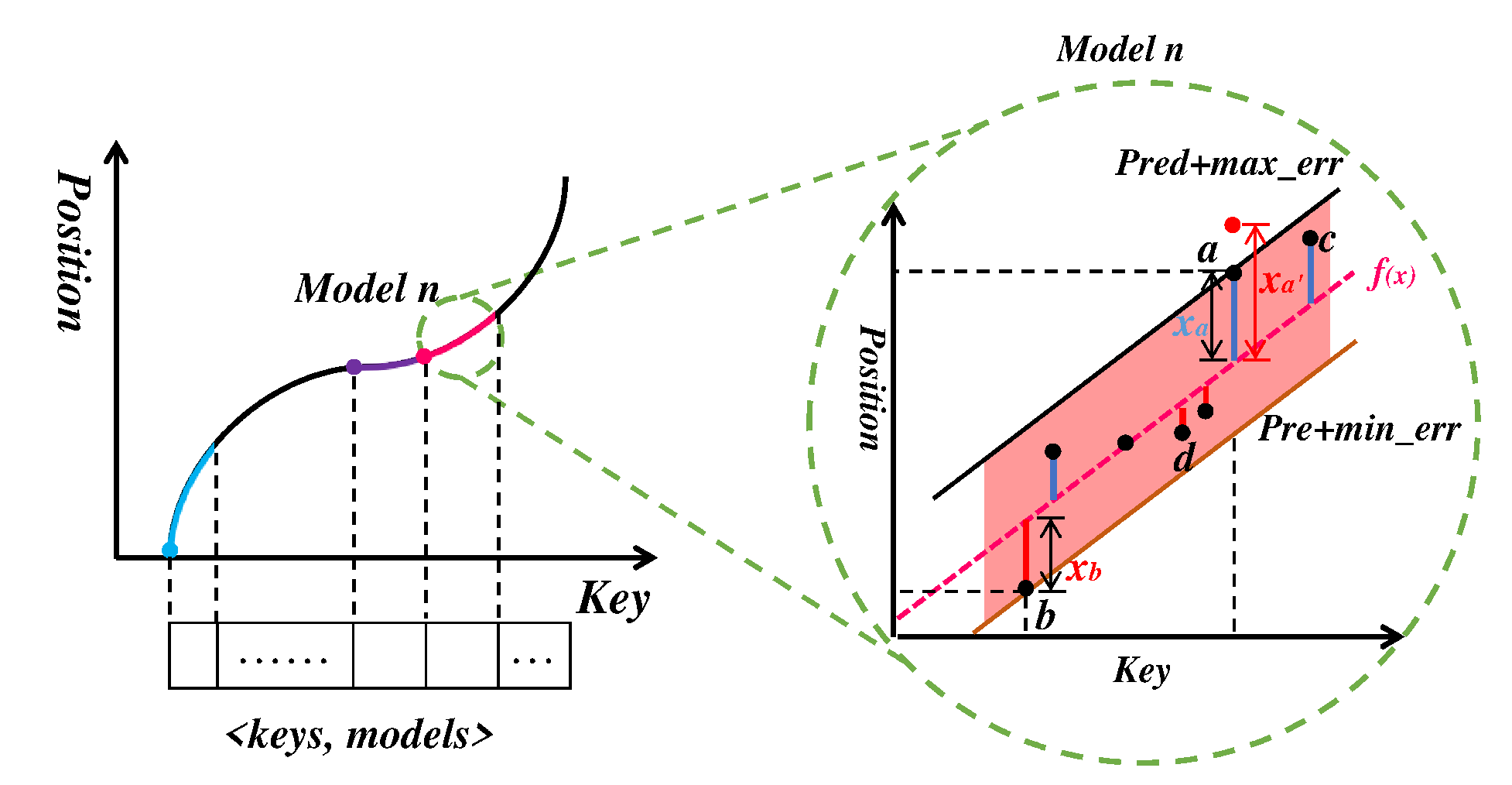}
	\caption{The case of failing to identifying data in insertion.}\label{fig:guarantee}
\end{figure}

AIDEL is constructed through LPA and the structure is shown of the left part in Figure \ref{fig:guarantee}. All the linear regression models are stored in the form of key-value, where the key is the first data covered by the model and the value is a pointer to the model.

\subsection{Scalability}\label{section:scalability}
Traditional B$^+$-tree achieves scalability by rebalancing the tree, which can insert new data while efficiently meeting range requests over sorted data. However, the scalability is not easy for learned indexes since inserting new data may incur an error that some data can't be found. Because the newly inserted data will change the positions of some data to keep all data in order, leading some data to exceed the prediction granularity. As shown in Figure \ref{fig:guarantee}, the red line represents one of the linear regression models generated by LPA, and the black points are the data covered by this model. Since $min\_err$ and $max\_err$ of the model are calculated via Equation \ref{equation:calculate_error} as described in Section \ref{section:learned_indexes}, the error $x_{a}$ of point $a$ meets the condition:
$$min\_err\leq x_{a}\leq max\_err$$
Therefore, point $a$ must be found by the model since the true position of $a$ meet the condition:
$$a\in [pred\_a+min\_err, pred\_a+max\_err]$$
where $pred\_a$ represents the prediction result of the linear model. Obviously, all the covered points can be found by the model, which however is not true when there are some newly inserted data. For example, we have to move point $a$ to $a'$ if we directly insert a new data before point $a$, which leads to the error:$$a'\notin [pred\_a'+min\_err, pred\_a'+max\_err]$$
since the new error $x_{a}'>max\_err$.

Learned indexes~\cite{kraska2018case} argue to build a delta-index~\cite{severance1976differential} to handle new inserts, which has been widely used in other structures such as Bigtable~\cite{chang2008bigtable} and A-tree~\cite{galakatos2018tree}. However, the design of delta-index incurs additional issues. We have to retrain the entire structure when delta-index is full. Moreover, range requests are inefficient, because the data are not in order due to being stored in delta-index and learned indexes respectively.

\begin{figure}[tbp]
	\centering\includegraphics[width=0.45\textwidth]{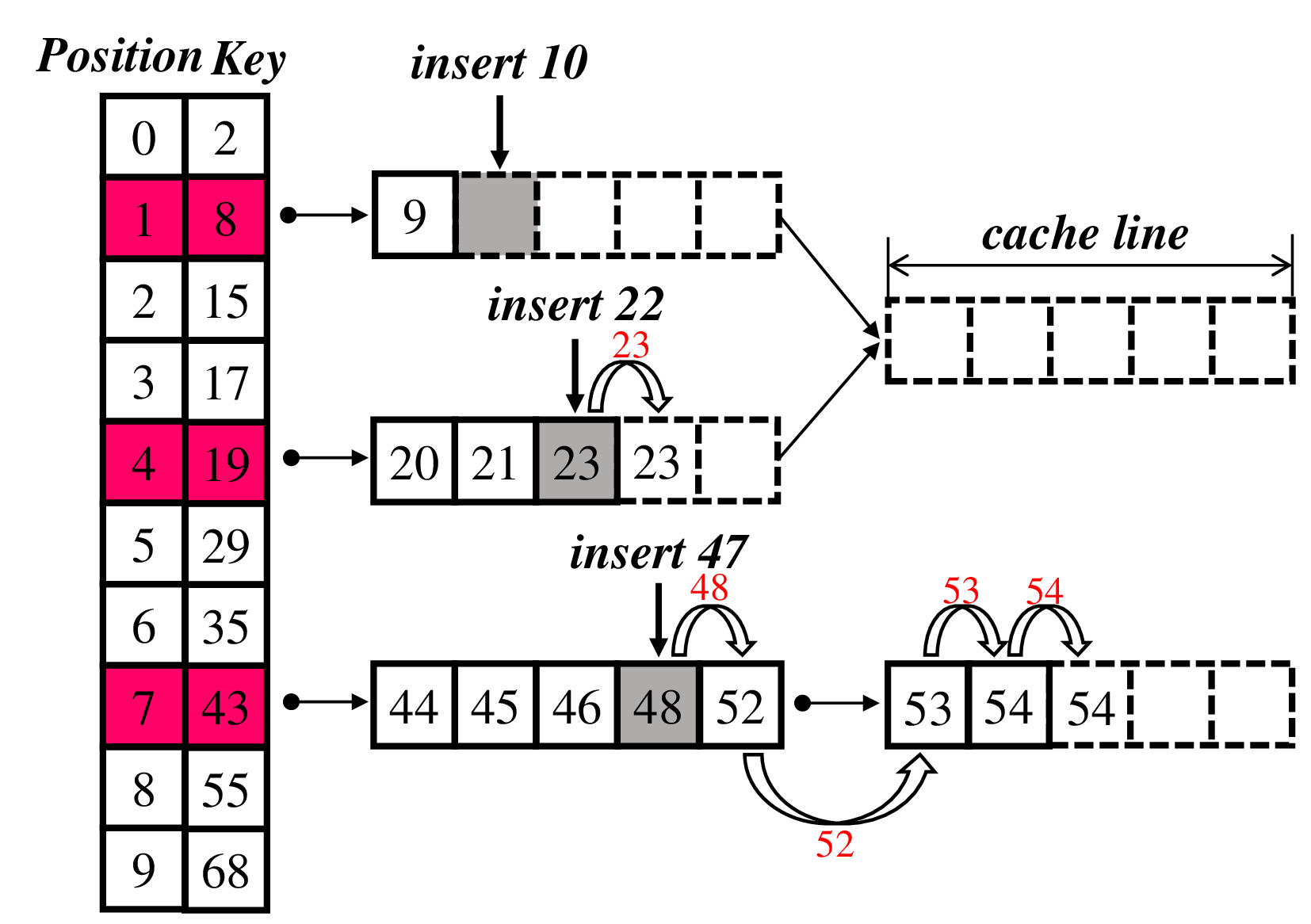}
	\caption{The insertion process of AIDEL.}\label{fig:scalability}
\end{figure}

The design goal of our scalable AIDEL structure is to avoid the errors that some data can't be found, while efficiently meeting range requests over sorted data. AIDEL achieves scalability through a structure of sorted lists which are appended behind the existing data. Figure \ref{fig:scalability} illustrates the insertion process. The data in $<key, position>$ represent the \textbf{\textit{existing data}}, while the data in the sorted lists are the newly inserted data. The insertion process can be divided into two steps: (1)Find a position in the prediction granularity $[pred+min\_err, pred+max\_err]$ which is given by the linear regression model. AIDEL will return a position whose key is first smaller than the new data if the prediction granularity doesn't contain this key. (2)Insert the data into sorted lists behind this position. For example, AIDEL returns the position 4 in the first step when we insert 22, since 19 is first smaller than 22 in the existing data. Then, 22 is inserted into the sorted list behind the data 19. To access data efficiently, the length of each list is aligned with a cache line to leverage the cache. AIDEL will assign a new sorted list when there are no empty positions in existing lists, as shown in Figure \ref{fig:scalability}.

It is worth noting that such designs meet two criteria of the design goal. First, the insertion process in sorted lists doesn't change the positions of existing data, which avoids the movements of existing data and hence guarantees that the data can always be found. Second, existing data and the inserted data in sorted lists are kept in order, which efficiently meets range requests.

\subsection{Retraining}\label{section:retraining}
Although AIDEL achieves scalability through the sorted lists, the performance will decrease if the sorted lists are too long. Because AIDEL has to spend a lot of time searching on sorted lists. One way to improve the performance is to retrain AIDEL. There are two types of retraining in AIDEL. One is to retrain all models and the other is to retrain part of the models.

The process of retraining all models can be divided into three steps. (1)Put all data into $<key, position>$ pairs and keep them in order, which are served as new existing data for training new models. This step is easy to be achieved since all data have been kept in order via sorted lists as described in Section \ref{section:scalability}. What we need to do is just to put the data from sorted lists after the corresponding existing data. For example, data 9 and 10 in the sorted list can be inserted after 8 without re-sorting as shown in Figure \ref{fig:scalability}. (2)Retrain the models via LPA algorithm on the new existing data. (3)Construct key-value pairs for the first stage to indicate how to choose each model, where the key is the first data covered by each model and the value is a pointer to the model.

One advantage of retraining in AIDEL is to retrain any models without affecting the whole structure, since all the models are independent. The process of retraining part of the models can also be divided into three steps. (1)Different from retraining all models, we only need to put the data that are covered by the model to be retrained into $<key, position>$ pairs. In Figure \ref{fig:scalability}, suppose the model to be retrained covers data 2, 8, 15, 17, 19, and we only need to put the data from two covered sorted lists (i.e., 9, 10 and 20, 21, 22, 23) into $<key, position>$ pairs. (2)Retrain the model that covers these data via LPA algorithm. LPA may generate multiple models if the error of obtained model is larger than the predefined $threshold$. (3) Update the information of retrained models to the first stage.

Compared with learned indexes, retraining AIDEL is more cost-efficient. First, learned indexes need to re-sort all data since the new data are stored separately from the existing ones and these data are not in order. However, AIDEL doesn't need to spend time on re-sorting since AIDEL guarantees that all data have been kept in order via sorted lists as described in Section \ref{section:scalability}. Second, AIDEL can retrain any model without affecting the whole structure. Learned indexes have to retrain the entire structure even if only one model needs to be retrained. Because each model in learned indexes is selected by another model according to Equation \ref{equation:calculation}, all models are highly dependent. In AIDEL, all models are independent since they are selected according to key-value pairs in the first stage. We can modify any models by updating the key-value pairs.

Although the training for ML models is usually considered to be time consuming, learned indexes indicate that the training doesn't consume much longer than a few seconds for 200M records with simple RMI~\cite{kraska2018case}. The core components in AIDEL are linear regression models which are simpler than learned indexes. Moreover, the training for ML models can be accelerated by powerful hardware such as GPUs and TPUs as describe in Section \ref{section:competitive_performance}.

\subsection{Data Partition}\label{section:paging}
In real-world applications, it is common to partition the data into different blocks and store them in separate regions, such as disks and distributed systems. However, learned indexes only consider the case where all data are stored in one contiguous block. The RMI structure in learned indexes is unsuitable to partition the data into different regions since the models are dependent. The entire structure of learned indexes needs to be retrained even if only one model needs to be modified as analyzed in Section \ref{section:retraining}.

Learned indexes outline several options to overcome this issue. First, it is possible to figure out the regions that are overlapped by multiple models through RMI and duplicate these data. Second, we can create an additional translation table for the conversion of different addresses. However, both methods are complicated and difficult to achieve the scalability. Moreover, all components in RMI have to be rebuilt once retraining is required.

AIDEL is easy to partition and store the data in different regions, because all the models are independent and the data are non-overlapped by each model. AIDEL can update (i.e., add, delete and change) each model by simply updating the $<key, model>$ pairs in the first stage. During the partitioning, AIDEL only needs to remove the corresponding models which cover the partition data, and train new models on these data. Interestingly, training on these data is particularly easy because the original models have already divided the data according to the data distribution. What we need to do is to train new models on each divided part.

%% file: experiment.tex
\section{Performance Evaluation}\label{section:experiment}
In this section, we evaluate the performance of AIDEL and compare it with state-of-the-art index structures including B$^+$-tree~\cite{knuth1997art}, FAST~\cite{kim2010fast} and learned indexes~\cite{kraska2018case}.

As the baseline, we use a popular B$^+$-tree implementation (stx::tree~\cite{stx::btree} version 0.9). In the B$^+$-tree, we use a fan-out of 128 since B$^+$-tree provides the best lookup performance with this configuration as described in ~\cite{kraska2018case}. We use the same configuration (i.e. a page size of 128) for learned indexes and AIDEL. FAST~\cite{kim2010fast} is the state-of-the-art SIMD optimized B$^+$-tree. However, the structure in FAST requires the tree size to be the power of 2, thus leading to larger space overhead. Additionally, it is difficult to update indexes in FAST because this structure has been optimized for cache and SIMD instructions. Learned indexes~\cite{kraska2018case} use ML models to address the index issues. Since no public implementations of learned indexes~\cite{kraska2018case} are available, we implement a 2-stage recursive model index with the same configuration as learned indexes. In order to compare the insertion performance, we build a delta-index to handle new inserts for learned indexes. Moreover, different configurations of the second stage in learned indexes, including 10K, 50K, 100K, 200K, are denoted as LI\_10K, LI\_50K, LI\_100K, LI\_200K, respectively.


We use several datasets with different distributions to evaluate the performance of index structures. Among them, we choose 2 real-world datasets (1)Weblogs, (2)DocId, and 1 synthetic dataset (3)Lognormal.

\begin{itemize}
\setlength{\itemsep}{0pt}
\setlength{\parsep}{0pt}
\setlength{\parskip}{0pt}
    \item {\bf Weblogs} dataset contains 200 million log entries and we use the timestamps as the indexes.
    \item {\bf DocId} contains five text collections in the form of bags-of-words, which has nearly 10 million instances in total. The DocID and WordID are used to identify unique documents and words, which are all non-linearly continuous.
    \item {\bf Lognormal} dataset is generated similarly with learned indexes, which contains 190 million unique values that follow the lognormal distribution with $\mu$=0 and $\sigma$=2, and each value is scaled up to be an integer up 1B.
\end{itemize}

\begin{figure}[tbp]
	\centering\includegraphics[width=3.2in]{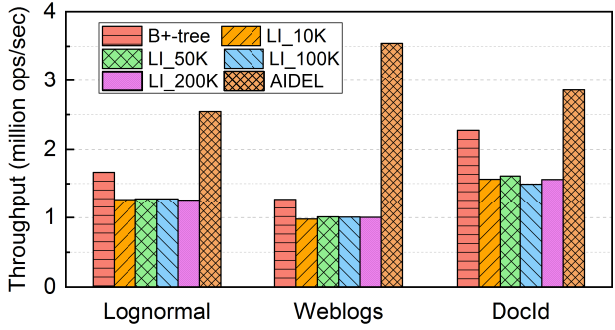}
	\caption{Insertion throughput in different datasets.}\label{fig:InsertionThroughput}
\end{figure}


We run experiments on a server that is equipped with an Intel 2.8 GHz 16-core CPU, 16 GB DRAM and 500 GB hard disk. The L1 and L2 caches of the CPU are 32KB and 256KB, respectively. The prototypes are developed under the Linux kernel 2.6.18 environment and we compile all implementations using g++ 8.1.0 with -O3 option.

\subsection{Insertion Performance}\label{section:insertion Performance}

In the experiments for measuring insertion performance, we compare AIDEL with B$^+$-tree and learned indexes with different configurations. Since there are no insertion functions in learned indexes, we build a delta-index to handle the new inserts. We don't compare AIDEL with FAST, because FAST can't handle new inserts unless reconstruct the entire structure.

Unlike the traditional index structures, both learned indexes and AIDEL based on ML models, require offline training, and we choose 10\% of the total data to execute the training in the experiments. Then we disorder the data to eliminate the impact of cache and the results of insertion throughput are shown in Figure \ref{fig:InsertionThroughput}. AIDEL improves the insertion throughput by 1.3$\times$ to 2.7$\times$ compared with the traditional B$^+$-trees. The insertion performance of learned indexes is low, because learned indexes have to check the learned ML models and delta-buffer to confirm they don't contain the data before insertion. We also observe that increasing the number of models is not useful for improving the insertion throughput, because the bottleneck of insertion is the delta-buffer instead of ML models. AIDEL has high insertion performance because after learning the patterns, AIDEL can quickly locate the approximate location of the new data and append the data into the sorted lists.

\begin{figure}[tbp]
	\centering\includegraphics[width=3.2in]{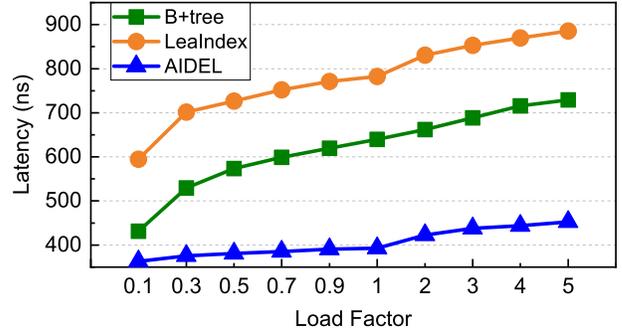}
	\caption{Insertion latency with different load factors.}\label{fig:InsertLatency}
\end{figure}

As shown in Figure \ref{fig:InsertLatency}, we evaluate the insertion latency of different index structures with the increase of the load factors. We define the load factor as the ratio of the inserted data to the training data. We observe that the insertion speed of AIDEL is the fastest, and the main reason is that AIDEL can quickly find the corresponding sorted lists after learning the patterns. If the newly inserted data follow the same patterns, AIDEL is most likely to deliver the data to each sorted list. However, the sorted lists may become longer as the inserted data increase, which will affect the insertion and query performance as described in Section \ref{section:scalability}. We then need to retrain the learned structures.

\begin{figure*}[htp]
	\centering
	\subfigure[Lookup throughput with no inserts]{
		\includegraphics[width=0.31\textwidth]{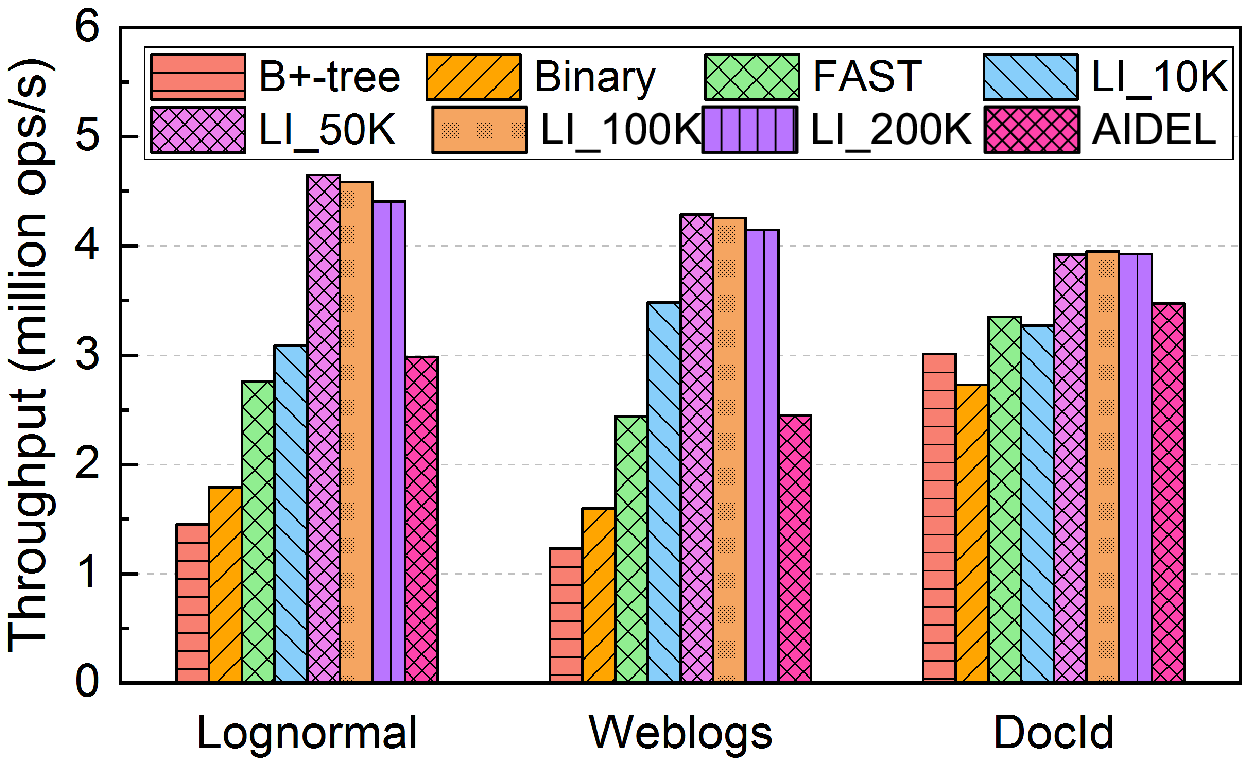}
	}
	\subfigure[Lookup throughput with the insert factor of 0.5]{
		\includegraphics[width=.31\textwidth]{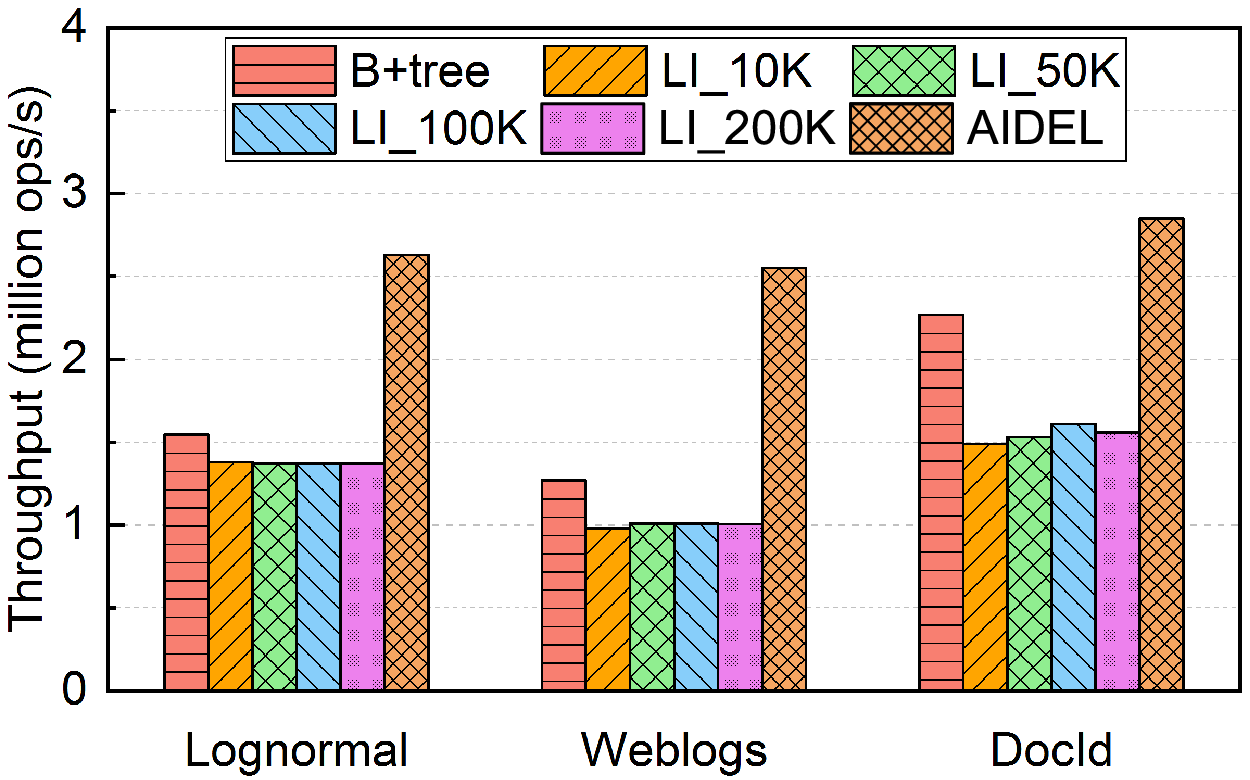}
	}
	\subfigure[Lookup throughput with the insert factor of 1]{
		\includegraphics[width=.31\textwidth]{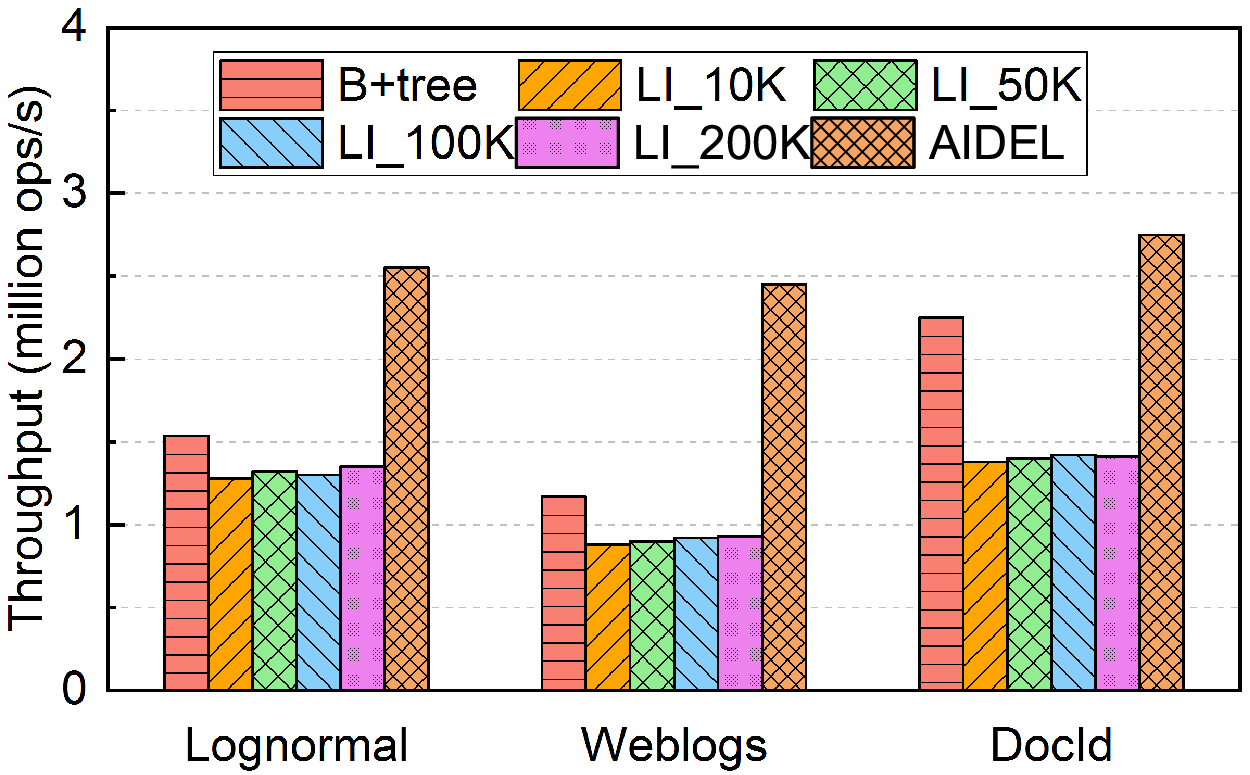}
	}
	\caption{Lookup throughput in different situations.}
	\label{fig:LookupThroughput}
\end{figure*}

\subsection{Lookup Performance}\label{section:lookup}

In the evaluation of lookup performance, we first insert the workloads into each index structure and then generate a new random workload which contains existing keys and non-existing keys to evaluate the lookup performance. We compare AIDEL against binary search, B$^+$-tree, FAST and learned indexes which configure different amounts of models in the second stage (i.e. 10K, 50K, 100K and 200K). We included binary search since this method represents the worst case where the prediction granularity is equal to the size of the whole dataset. We also compare the lookup performance of the index structures with and without insertion operations. As shown in Section \ref{section:insertion Performance}, we build a delta-index to achieve the scalability for learned indexes.

We evaluate the lookup performance of these different index structures with no inserts and the results are shown in Figure \ref{fig:LookupThroughput}(a). When there are no inserts, learned indexes improve the lookup performance by 1.3$\times$ to 3.1$\times$ compared with the B$^+$-tree, because learned indexes can quickly find the corresponding model (i.e. the model knows where the look-up key may locate) through the recursive model index (RMI). AIDEL improves the lookup performance by 1.2$\times$ to 2.1$\times$ compared with the B$^+$-tree, since the models in AIDEL are independent, which have to leverage the traditional methods to find the corresponding model. The lookup performance of AIDEL is nearly the same as FAST and is comparable to learned indexes without inserts.

We further add some inserts to evaluate the lookup performance (e.g. load factors of 0.5 and 1) and the results are shown in Figures \ref{fig:LookupThroughput}(b) and (c). We don't include FAST since FAST can't handle inserts unless reconstruct the entire structure. Compared with the experimental results in Figure \ref{fig:LookupThroughput}, learned indexes can't achieve the same lookup performance as no inserts. The main reason is that learned indexes handle the inserts with a delta-index and have to lookup both structures in one lookup operation. In contrast, AIDEL improves the lookup performance by 1.3$\times$ to 2.1$\times$ when inserting new data, since AIDEL achieves scalability through the sorted lists and can deliver the data to each sorted list. Moreover, the independent structures make retraining easy as described in Section \ref{section:retraining}.

\subsection{Memory Overhead}
Compared with the traditional index structures, learned structures consume less memory. B$^+$-tree stores all the data in the nodes while learned structures manage the data by ML models. In the learned indexes, the number of models is configured by the user in advance. Memory overhead can be calculated according to this configuration. However, AIDEL uses the learning probe algorithm to adaptively assign different ML models to learn the data distribution, and the number of the ML models is to be known after learning.

\begin{table*}[htp]
	\centering
	\caption{The numbers of models in learned indexes and AIDEL on the dataset of lognormal.}
	\label{table:Model_Numbers}
	\begin{tabular}{p{3cm}<{\centering}|p{2cm}<{\centering}|p{1.5cm}<{\centering}|p{1.5cm}<{\centering}|p{1.5cm}<{\centering}|p{1.5cm}<{\centering}|p{2cm}<{\centering}}
		\hline
		\multirow{2}*{{\bf Threshold}} & \multirow{2}*{{\bf Type}} &
		\multicolumn{4}{@{}c|}{{\bf Learned Indexes}} & \multirow{2}*{{\bf AIDEL}} \\ \cline{3-6} & & {\bf 10K} & {\bf 50K} & {\bf 100K} & {\bf 200K} \\ \hline \hline
		\multirow{2}*{32} & total & 10,000 & 50,000 & 100,000 & 200,000 & 58,695 \\
		\cline{2-7} & unsatisfied & 9,934 & 23,569 & 19,510 & 9,287 & 0 \\ \hline
		\multirow{2}*{64} & total & 10,000 & 50,000 & 100,000 & 200,000 & 15,301 \\
		\cline{2-7} & unsatisfied & 5,905 & 2,467 & 396 & 17 & 0 \\ \hline
		\multirow{2}*{128} & total & 10,000 & 50,000 & 100,000 & 200,000 & 4,132 \\
		\cline{2-7} & unsatisfied & 896 & 8 & 0 & 0 & 0 \\ \hline
		\multirow{2}*{256} & total & 10,000 & 50,000 & 100,000 & 200,000 & 991 \\
		\cline{2-7} & unsatisfied & 13 & 0 & 0 & 0 & 0 \\ \hline
	\end{tabular}
\end{table*}

\begin{figure}[tbp]
	\centering\includegraphics[width=3in]{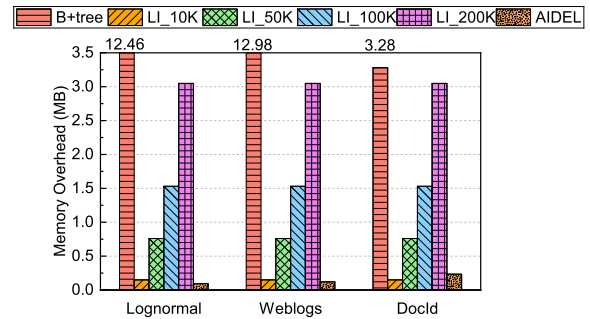}
	\caption{Memory overheads.}\label{fig:MemoryOverhead}
\end{figure}

We evaluate the metadata overhead of each index structure, i.e., the intermediate nodes in the B$^+$-tree and the ML models in learned structures. The experimental results are shown in Figure \ref{fig:MemoryOverhead}. Both learned indexes and AIDEL consume less memory than B$^+$-tree by 14$\times$ to 130$\times$. The main reason is that one trained ML model can cover lots of data, and we only need to store the parameters of the trained ML model. In the context of our paper, all models are linear regression models, which only contain two parameters (i.e., the slop and intercept). Moreover, AIDEL can save more memory than the leaned indexes, since the models in AIDEL are trained according to the data distribution.

\subsection{Model Numbers}\label{section:model_numbers}
Compared with learned indexes, AIDEL eliminates the invalid and redundant models by using the learning probe algorithm, which uses less models than learned indexes. We evaluate the used models with different $threshold$ values, where the $threshold$ is the max prediction error that we can tolerate.

The number of models in learned indexes can be determined in advance according to the configurations, and the main components in learned indexes are the models in the second stage whose number is 10K, 50K, 100K, 200K, respectively. However, the models whose $errors$ are larger than $threshold$ can't be used, since these trained models are not able to offer a small enough range, which contains the look-up key. We have to use the traditional index structures such as the B$^+$-tree to replace the invalid models as described in learned indexes~\cite{kraska2018case}.

We evaluate the numbers of the invalid models in the experiments, and the results that evaluates on Lognormal are shown in Table \ref{table:Model_Numbers}. AIDEL can use fewer models than learned indexes under the same $threshold$, since the models in AIDEL can cover the data that have the same patterns as many as possible according to the data distribution. Moreover, the models in AIDEL are all valid (i.e. $error<=threshold$), since the learning probe algorithm guarantees that only the models meet the condition $error<=threshold$ can be appended to AIDEL as described in Section \ref{section:lepa_algorithm}.

However, learned indexes use the strategy of normalization to partition the data. Once we fail to configure sufficient models to learn the data, lots of trained models are invalid. In contrast, if the learned indexes use sufficient models, e.g., 200K, to learn the data, most of the models are redundant since learned indexes can't assign models according to the data distribution as analyzed in Section \ref{section:strategy_comparison}. For example, from Table \ref{table:Model_Numbers}, the learned indexes use 200K models to allow most models to be valid, but there still exists 17 invalid models with the $threshold$ of 64. However, AIDEL only needs nearly 15K models to learn the data distribution and guarantees that all these models are valid, since our proposed learning probe algorithm ensures that only the model whose prediction error is smaller than the predefined threshold can be appended into AIDEL. The main reason is that AIDEL can adaptively assign different models according to the data distribution.

%% file: related_work.tex
\section{Related Work}
B$^+$-trees~\cite{comer1979ubiquitous} are designed to accelerate searches on disk-based database systems and different variants have been proposed over the past few decades~\cite{graefe2001b}. B$^+$-trees~\cite{bayer2002organization} are used for disk based systems and T-trees~\cite{lehman1986study} are proposed to be a replacement since the main memory sizes become large enough to store entire database. In order to provide faster query performance in the database, several cache conscious B$^+$-tree variants are proposed since Rao et al.~\cite{rao1999cache} showed that B$^+$-trees have good cache behavior on modern processors. They propose CSS-tree~\cite{rao1999cache} and CSB+-tree~\cite{rao2000making} to efficiently support index operations by exploiting the cache. More recently, with powerful hardware, there are some schemes aiming to provide higher performance by using SIMD instructions such as FAST~\cite{kim2010fast} or GPUs~\cite{kaczmarski2012b+,kim2010fast,shahvarani2016hybrid}.

The above methods mainly focus on the lookup time of the B$^+$-trees while overlooking the memory overhead. However, B$^+$-tree often consumes much storage space. There are several schemes on compressing indexes to reduce the size of keys via prefix/suffix truncation, dictionary compression and key normalization~\cite{goldstein1998compressing,boehm2011efficient,rao1999cache,neumann2008rdf}, or hybrid hot/cold indexes~\cite{zhang2016reducing}, which use a two-stage index to reduce the memory overhead. Learned indexes~\cite{kraska2018case} present a different way to compress indexes, which depend on the data distribution and achieve orders-of-magnitude less storage consumption compared with traditional B$^+$-trees.

Approximate indexes are used to reduce the memory overhead, such as BF-tree~\cite{athanassoulis2014bf}, A-tree~\cite{galakatos2018tree} and Learned indexes~\cite{kraska2018case}. BF-tree and A-tree use a B$^+$-tree to store information about a region of the dataset, but the leaf nodes in a BF-tree are Bloom filters while in an A-tree are linear segments. Learned indexes\cite{kraska2018case} proposed recursive model index to narrow the search range of a record, by using a simple neural network in the first stage, while many linear models in the second stage. Additionally, BF-trees fail to consider data distribution, while A-tree, learned indexes and AIDEL all exploit the properties about the data distribution.

Since there are many researches on machine learning accelerators including GPUs, FPGAs, ASICs, PIMs and NVMs~\cite{chen2014diannao,chen2016eyeriss,kim2016neurocube,wang2017towards,chi2016prime,shafiee2016isaac}, we can use more advanced ML models such as convolutional neural network and the mixture of experts~\cite{krizhevsky2012imagenet,simonyan2014very,shazeer2017outrageously} for learned indexes. Moreover, the relationship between the keys and positions is similar to the CDF, and there has many researches on estimating the distribution of data~\cite{dvoretzky1956asymptotic,magdon1999neural,huang2011cumulative}.

Unlike them, the design goal of our paper is not to completely replace the traditional B$^+$-trees, but to complement the existing schemes. At the same time, the scheme in this paper also needs to use traditional index structures to store the linear regression models and thus all these related techniques are orthogonal to AIDEL.

%% file: secpm-arXiv.bbl
\begin{thebibliography}{10}

\bibitem{alexiou2013adaptive}
{\sc Alexiou, K., Kossmann, D., and Larson, P.-{\AA}.}
\newblock Adaptive range filters for cold data: Avoiding trips to siberia.
\newblock {\em Proceedings of the VLDB Endowment 6}, 14 (2013), 1714--1725.

\bibitem{athanassoulis2014bf}
{\sc Athanassoulis, M., and Ailamaki, A.}
\newblock Bf-tree: approximate tree indexing.
\newblock {\em Proceedings of the VLDB Endowment 7}, 14 (2014), 1881--1892.

\bibitem{bayer2002organization}
{\sc Bayer, R., and McCreight, E.}
\newblock Organization and maintenance of large ordered indexes.
\newblock In {\em Software pioneers}. Springer, 2002, pp.~245--262.

\bibitem{bayer1977prefix}
{\sc Bayer, R., and Unterauer, K.}
\newblock Prefix b-trees.
\newblock {\em ACM Transactions on Database Systems (TODS) 2}, 1 (1977),
  11--26.

\bibitem{stx::btree}
{\sc Bingmann, T.}
\newblock Stx b+ tree c++ template classes, 2007.

\bibitem{boehm2011efficient}
{\sc Boehm, M., Schlegel, B., Volk, P.~B., Fischer, U., Habich, D., and Lehner,
  W.}
\newblock Efficient in-memory indexing with generalized prefix trees.
\newblock In {\em BTW\/} (2011), vol.~180, pp.~227--246.

\bibitem{chang2008bigtable}
{\sc Chang, F., Dean, J., Ghemawat, S., Hsieh, W.~C., Wallach, D.~A., Burrows,
  M., Chandra, T., Fikes, A., and Gruber, R.~E.}
\newblock Bigtable: A distributed storage system for structured data.
\newblock {\em ACM Transactions on Computer Systems (TOCS) 26}, 2 (2008), 4.

\bibitem{chang2000b}
{\sc Chang, Y.-C., Chang, Y.-W., Wu, G.-M., and Wu, S.-W.}
\newblock B*-trees: a new representation for non-slicing floorplans.
\newblock In {\em Proceedings of the 37th Annual Design Automation
  Conference\/} (2000), ACM, pp.~458--463.

\bibitem{chen2014diannao}
{\sc Chen, T., Du, Z., Sun, N., Wang, J., Wu, C., Chen, Y., and Temam, O.}
\newblock Diannao: A small-footprint high-throughput accelerator for ubiquitous
  machine-learning.
\newblock {\em ACM Sigplan Notices 49}, 4 (2014), 269--284.

\bibitem{chen2016eyeriss}
{\sc Chen, Y.-H., Emer, J., and Sze, V.}
\newblock Eyeriss: A spatial architecture for energy-efficient dataflow for
  convolutional neural networks.
\newblock In {\em ACM SIGARCH Computer Architecture News\/} (2016), vol.~44,
  IEEE Press, pp.~367--379.

\bibitem{chi2016prime}
{\sc Chi, P., Li, S., Xu, C., Zhang, T., Zhao, J., Liu, Y., Wang, Y., and Xie,
  Y.}
\newblock Prime: A novel processing-in-memory architecture for neural network
  computation in reram-based main memory.
\newblock In {\em ACM SIGARCH Computer Architecture News\/} (2016), vol.~44,
  IEEE Press, pp.~27--39.

\bibitem{comer1979ubiquitous}
{\sc Comer, D.}
\newblock Ubiquitous b-tree.
\newblock {\em ACM Computing Surveys (CSUR) 11}, 2 (1979), 121--137.

\bibitem{dvoretzky1956asymptotic}
{\sc Dvoretzky, A., Kiefer, J., and Wolfowitz, J.}
\newblock Asymptotic minimax character of the sample distribution function and
  of the classical multinomial estimator.
\newblock {\em The Annals of Mathematical Statistics\/} (1956), 642--669.

\bibitem{fan2014cuckoo}
{\sc Fan, B., Andersen, D.~G., Kaminsky, M., and Mitzenmacher, M.~D.}
\newblock Cuckoo filter: Practically better than bloom.
\newblock In {\em Proceedings of the 10th ACM International on Conference on
  emerging Networking Experiments and Technologies\/} (2014), ACM, pp.~75--88.

\bibitem{galakatos2018tree}
{\sc Galakatos, A., Markovitch, M., Binnig, C., Fonseca, R., and Kraska, T.}
\newblock A-tree: A bounded approximate index structure.
\newblock {\em arXiv preprint arXiv:1801.10207\/} (2018).

\bibitem{goldstein1998compressing}
{\sc Goldstein, J., Ramakrishnan, R., and Shaft, U.}
\newblock Compressing relations and indexes.
\newblock In {\em Data Engineering, 1998. Proceedings., 14th International
  Conference on\/} (1998), IEEE, pp.~370--379.

\bibitem{graefe2001b}
{\sc Graefe, G., and Larson, P.-A.}
\newblock B-tree indexes and cpu caches.
\newblock In {\em Data Engineering, 2001. Proceedings. 17th International
  Conference on\/} (2001), IEEE, pp.~349--358.

\bibitem{huang2011cumulative}
{\sc Huang, J.~C., and Frey, B.~J.}
\newblock Cumulative distribution networks and the derivative-sum-product
  algorithm: Models and inference for cumulative distribution functions on
  graphs.
\newblock {\em Journal of Machine Learning Research 12}, Jan (2011), 301--348.

\bibitem{hwang2018endurable}
{\sc Hwang, D., Kim, W.-H., Won, Y., and Nam, B.}
\newblock Endurable transient inconsistency in byte-addressable persistent
  b+-tree.
\newblock In {\em 16th USENIX Conference on File and Storage Technologies\/}
  (2018), p.~187.

\bibitem{kaczmarski2012b+}
{\sc Kaczmarski, K.}
\newblock B+-tree optimized for gpgpu.
\newblock In {\em OTM Confederated International Conferences" On the Move to
  Meaningful Internet Systems"\/} (2012), Springer, pp.~843--854.

\bibitem{kim2010fast}
{\sc Kim, C., Chhugani, J., Satish, N., Sedlar, E., Nguyen, A.~D., Kaldewey,
  T., Lee, V.~W., Brandt, S.~A., and Dubey, P.}
\newblock Fast: fast architecture sensitive tree search on modern cpus and
  gpus.
\newblock In {\em Proceedings of the 2010 ACM SIGMOD International Conference
  on Management of data\/} (2010), ACM, pp.~339--350.

\bibitem{kim2016neurocube}
{\sc Kim, D., Kung, J., Chai, S., Yalamanchili, S., and Mukhopadhyay, S.}
\newblock Neurocube: A programmable digital neuromorphic architecture with
  high-density 3d memory.
\newblock In {\em Computer Architecture (ISCA), 2016 ACM/IEEE 43rd Annual
  International Symposium on\/} (2016), IEEE, pp.~380--392.

\bibitem{knuth1997art}
{\sc Knuth, D.~E.}
\newblock {\em The art of computer programming: sorting and searching}, vol.~3.
\newblock Pearson Education, 1997.

\bibitem{kraska2018case}
{\sc Kraska, T., Beutel, A., Chi, E.~H., Dean, J., and Polyzotis, N.}
\newblock The case for learned index structures.
\newblock In {\em Proceedings of the 2018 International Conference on
  Management of Data\/} (2018), ACM, pp.~489--504.

\bibitem{krizhevsky2012imagenet}
{\sc Krizhevsky, A., Sutskever, I., and Hinton, G.~E.}
\newblock Imagenet classification with deep convolutional neural networks.
\newblock In {\em Advances in neural information processing systems\/} (2012),
  pp.~1097--1105.

\bibitem{lehman1986study}
{\sc Lehman, T.~J., and Carey, M.~J.}
\newblock A study of index structures for main memory database management
  systems.
\newblock In {\em Proc. VLDB\/} (1986), vol.~1.

\bibitem{magdon1999neural}
{\sc Magdon-Ismail, M., and Atiya, A.~F.}
\newblock Neural networks for density estimation.
\newblock In {\em Advances in Neural Information Processing Systems\/} (1999),
  pp.~522--528.

\bibitem{neumann2008rdf}
{\sc Neumann, T., and Weikum, G.}
\newblock Rdf-3x: a risc-style engine for rdf.
\newblock {\em Proceedings of the VLDB Endowment 1}, 1 (2008), 647--659.

\bibitem{rao1999cache}
{\sc Rao, J., and Ross, K.~A.}
\newblock Cache conscious indexing for decision-support in main memory.
\newblock In {\em VLDB\/} (1999), vol.~99, Citeseer, pp.~78--89.

\bibitem{rao2000making}
{\sc Rao, J., and Ross, K.~A.}
\newblock Making b+-trees cache conscious in main memory.
\newblock In {\em Acm Sigmod Record\/} (2000), vol.~29, ACM, pp.~475--486.

\bibitem{richter2015seven}
{\sc Richter, S., Alvarez, V., and Dittrich, J.}
\newblock A seven-dimensional analysis of hashing methods and its implications
  on query processing.
\newblock {\em Proceedings of the VLDB Endowment 9}, 3 (2015), 96--107.

\bibitem{severance1976differential}
{\sc Severance, D.~G., and Lohman, G.~M.}
\newblock Differential files: their application to the maintenance of large
  databases.
\newblock {\em ACM Transactions on Database Systems (TODS) 1}, 3 (1976),
  256--267.

\bibitem{shafiee2016isaac}
{\sc Shafiee, A., Nag, A., Muralimanohar, N., Balasubramonian, R., Strachan,
  J.~P., Hu, M., Williams, R.~S., and Srikumar, V.}
\newblock Isaac: A convolutional neural network accelerator with in-situ analog
  arithmetic in crossbars.
\newblock {\em ACM SIGARCH Computer Architecture News 44}, 3 (2016), 14--26.

\bibitem{shahvarani2016hybrid}
{\sc Shahvarani, A., and Jacobsen, H.-A.}
\newblock A hybrid b+-tree as solution for in-memory indexing on cpu-gpu
  heterogeneous computing platforms.
\newblock In {\em Proceedings of the 2016 International Conference on
  Management of Data\/} (2016), ACM, pp.~1523--1538.

\bibitem{shazeer2017outrageously}
{\sc Shazeer, N., Mirhoseini, A., Maziarz, K., Davis, A., Le, Q., Hinton, G.,
  and Dean, J.}
\newblock Outrageously large neural networks: The sparsely-gated
  mixture-of-experts layer.
\newblock {\em arXiv preprint arXiv:1701.06538\/} (2017).

\bibitem{simonyan2014very}
{\sc Simonyan, K., and Zisserman, A.}
\newblock Very deep convolutional networks for large-scale image recognition.
\newblock {\em arXiv preprint arXiv:1409.1556\/} (2014).

\bibitem{wang2017towards}
{\sc Wang, Y., Zhang, M., and Yang, J.}
\newblock Towards memory-efficient processing-in-memory architecture for
  convolutional neural networks.
\newblock In {\em ACM SIGPLAN Notices\/} (2017), vol.~52, ACM, pp.~81--90.

\bibitem{wu2018wormhole}
{\sc Wu, X., Ni, F., and Jiang, S.}
\newblock Wormhole: A fast ordered index for in-memory data management.
\newblock {\em arXiv preprint arXiv:1805.02200\/} (2018).

\bibitem{zhang2016reducing}
{\sc Zhang, H., Andersen, D.~G., Pavlo, A., Kaminsky, M., Ma, L., and Shen, R.}
\newblock Reducing the storage overhead of main-memory oltp databases with
  hybrid indexes.
\newblock In {\em Proceedings of the 2016 International Conference on
  Management of Data\/} (2016), ACM, pp.~1567--1581.

\bibitem{zukowski2006super}
{\sc Zukowski, M., Heman, S., Nes, N., and Boncz, P.}
\newblock Super-scalar ram-cpu cache compression.
\newblock In {\em Data Engineering, 2006. ICDE'06. Proceedings of the 22nd
  International Conference on\/} (2006), IEEE, pp.~59--59.

\end{thebibliography}
